# Generation of Gamma-ray Beam with Orbital Angular Momentum in the QED Regime

Chen Liu, [1,2] Baifei Shen, [1,3,a] Xiaomei Zhang, [1, a] Yin Shi, [1] Liangliang Ji, [1] Wenpeng Wang, [1] Longqing Yi, [1] Lingang Zhang, [1,2] Tongjun Xu, [1,2] Zhikun Pei[1,2] and Zhizhan Xu[1]

[1]*State Key Laboratory of High Field Laser Physics, Shanghai Institute of Optics and Fine Mechanics, Chinese Academy of Sciences, P. O. Box 800-211, Shanghai 201800, China*

[2]*University of Chinese Academy of Sciences, Beijing 100049, China*

[3]*IFSA Collaborative Innovation Center, Shanghai Jiao Tong University, Shanghai 200240, China*

We propose a scheme to generate gamma-ray photons with an orbital angular momentum (OAM) and high energy simultaneously from laser-plasma interactions by irradiating a circularly polarized Laguerre–Gaussian laser on a thin plasma target. The spin angular momentum and OAM are first transferred to electrons from the driving laser photons, and then the OAM is transferred to the gamma-ray photons from the electrons through quantum radiation. This scheme has been demonstrated using three-dimensional quantum electrodynamics particle-in-cell simulation. The topological charge, chirality and carrier-envelope phase of the short ultra-intense vortex laser can be revealed according to the pattern feature of the energy density of radiated photons.

[a)] Authors to whom correspondence should be addressed: bfshen@mail.shcnc.ac.cn and zhxm@siom.ac.cn

Light beams with spiral phase front, also known as "twisted light," carry orbital angular momentum (OAM) in the propagation direction. In 1992, Allen *et al*. first demonstrated that Laguerre–Gaussian (LG) lasers possess this property [1]. Since then, the OAM of the LG laser has been extensively investigated for a wide range of applications in micromanipulation, quantum information, imaging and astrophysics [2-13]. Recently, an LG laser of relativistic intensity has been obtained by a scheme called "light fan" [14]. This scheme extends the research of the OAM into the relativistic regime, such as, in laser-driven particle acceleration [15,16], vortex high-order harmonics generation (HHG) [17], Raman Scattering [18] etc.

Owing to the wave-particle duality, matter waves with spiral phase fronts, or vortex particles, also possess an OAM. This is based on the similarity between the Schrödinger wave equation and Maxwell's equations. The generation of vortex particles such as electrons [19-21], neutrons [22,23], and photons of low energy (<< 1 MeV) [24-27] have been studied theoretically and experimentally. However, gamma-ray photons of high energy (> 1 MeV) carrying the OAM have not been realized. Recent multi-petawatt laser facilities [28] have made it possible for laser-plasma interaction in the quantum electrodynamics (QED) regime [29-32], where quantum radiation becomes dominant. In this regime, a gamma ray is generated in a random manner. It is incoherent and behaves more like particle beams. The transfer of the OAM from the LG laser to electrons has been verified [33], and it can be further transferred to the gamma-ray photons through the quantum radiation of energetic electrons in the QED regime, making it a promising scheme for generating gamma-ray photons with an OAM.

In this paper, we illustrate that gamma-ray photons with OAM can be generated from the interaction of an intense circularly polarized (CP) LG laser with a thin plasma target. This is confirmed by using three-dimensional QED particle-in-cell (PIC) simulations based on code EPOCH [34,35], which includes QED effects to deal with the quantum radiation. It has been found that the OAM of high energy gamma-ray photons is transferred from the spin angular momentum (SAM) and OAM of the driving laser. More importantly, the total energy density of the gamma-ray photons passing through a plane has a featured distribution. It consists of several parts and the total number is equal to the total angular momentum carried by each photon of the driving laser normalized to the Planck constant $\hbar$, or the number $l+s$, where $s$ is the chirality of the CP laser, $s=\pm 1$ corresponds to the right-hand and left-hand CP laser, respectively, and $l$ is the topological charge of the LG laser. This pattern also includes the information of the carrier-envelope phase (CEP) of the driving ultra-short laser. Therefore, the topological charge, chirality and CEP of the ultra-intense vortex laser pulse are expected to be detected by using the present scheme. And the generated high energy gamma photons may be also used in studying the related astrophysics phenomenon.

A right-hand CP LG laser is described as

$$a(\mathrm{LG}_{lp}) = (-1)^p a_0 (\frac{\sqrt{2}r}{r_0})^l \exp(\frac{-r^2}{r_0^2}) \exp(-il\phi) L_p^l (\frac{2r^2}{r_0^2}) \sin^2(\frac{\pi t}{2\tau_0})$$
$$\times [\sin(\omega t + \frac{\pi}{2} - \omega \tau_0 + \phi_{CEP})\vec{e}_y - \cos(\omega t + \frac{\pi}{2} - \omega \tau_0 + \phi_{CEP})\vec{e}_z].$$

For a photon of the CP LG laser, it carries the SAM of $s\hbar$ and OAM of $l\hbar$. In our simulation, the right-hand CP LG$_{10}$ mode is used as an example, which implies that $l=1, p=0, s=1$ and $L_0^1(\frac{2r^2}{r_0^2})=1$. The total angular momentum per photon of the driving laser beam on an average is $(l+s)\hbar = 2\hbar$. Here, $r_0 = 5\lambda$ is the radius of the laser spot, $\tau_0 = 5T$ is the pulse duration, $\lambda = 0.8\mu\mathrm{m}$ is the laser wavelength, $\phi_{CEP} = -\frac{\pi}{2}$ is the CEP, $T$ is the laser period, and $\phi$ is the azimuthal angle. The dimensionless laser amplitude $a_0$ is $300\sqrt{2}$, corresponding to a peak intensity of $I = 3.85 \times 10^{23} \mathrm{W/cm}^2$. The simulation box is $60\lambda(x) \times 60\lambda(y) \times 60\lambda(z)$ with $1200 \times 600 \times 600$ cells and 20 macro-particles per cell. The plasma occupies the $30\lambda < x < 31\lambda$ region in the laser propagation direction, and $-15\lambda < y(z) < 15\lambda$ in the transverse direction. The initial target density is $n_0 = 30 n_c = 5.16 \times 10^{22}\,\text{cm}^{-3}$, where $n_c = m_e \omega^2 / 4\pi e^2$ is the critical density, $m_e$ is the electron mass, $\omega$ is the laser frequency, and $e$ is the electron charge.

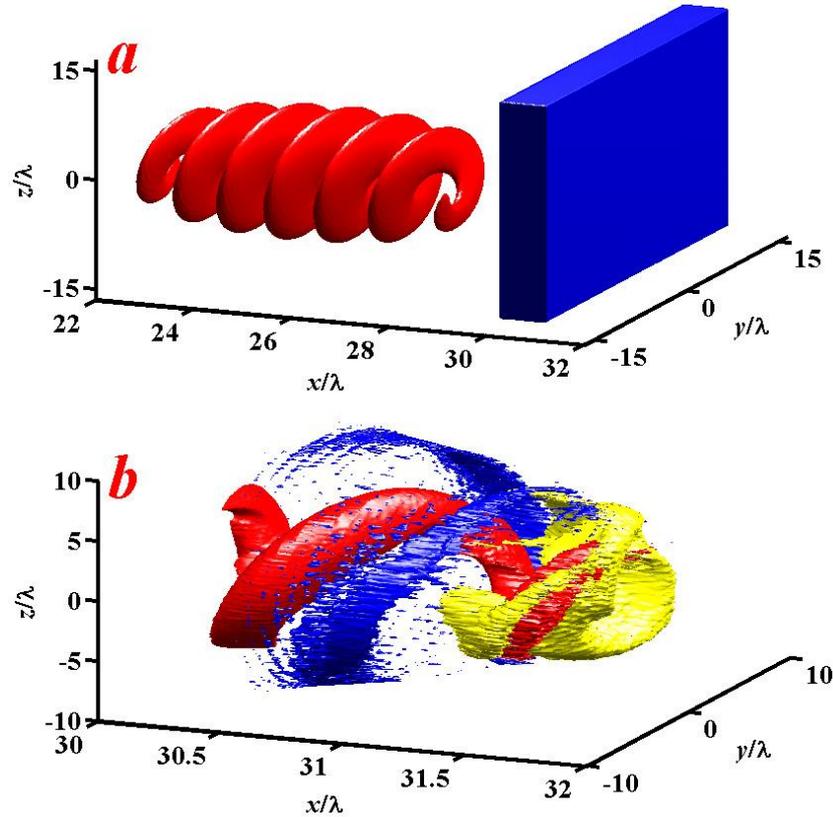

Fig. 1 Sketch of the generation of the gamma-ray photons with OAM from the laser-plasma interaction. (a) The initial setup of the CP $LG_{10}$ laser (red) and plasma target (blue) before the interaction. (b) Snapshot of the laser (red), electrons (blue) and generated gamma-ray photons (yellow) during the interaction.

The proposed scheme of the generation of the gamma-ray photons with OAM from the laser-plasma interaction is shown in Fig. 1. An intense CP $LG_{10}$ laser impinges normally on the thin hydrogen target and propagates through it. When the interaction begins, the CP $LG_{10}$ laser drives a bunch of electrons with an obvious vortex structure rotating around the $x$-axis, then these energetic electrons generate a forward burst of gamma-ray photons ahead of these electrons and simultaneously transfer OAM to them. Figure 1(b) shows a snapshot of this interaction.

In order to demonstrate that the generated gamma-ray photons carry OAM, the variation of OAM of different particles during the interaction is shown in Fig. 2(a). Here, OAM is calculated by $\text{OAM} = yp_z - zp_y$, $p_y, p_z$ are the linear momentum of the particles in $y$ and $z$ direction, respectively. It shows that the gamma-ray photons carrying OAM are generated from the laser-plasma interaction and the OAM of the gamma-ray photons increases as their number increases. At the end of the interaction, the OAM of the gamma-ray photons reaches $3.34 \times 10^{-14}$ kgm$^2$/s and compared to that of the driving laser ($2.33 \times 10^{-12}$ kgm$^2$/s), the conversion efficiency is 1.43%. In addition, the total energy of the generated gamma-ray photons is 86 J. That is to say, at least $N = 3.47 \times 10^{20}$ photons of the driving LG$_{10}$ laser pulse is consumed to generate the gamma-ray photons. Considering the OAM of the gamma-ray photons, when one photon of driving laser is consumed, the average OAM absorbed by a gamma-ray photon is approximate to $0.90\hbar$. Although the angular momentum carried by a single photon of the right-hand CP LG$_{10}$ laser is $2\hbar$ on an average, the angular momentum is not uniform along the radial direction $\vec{r}$, which can be expressed as $\text{OAM} = (l - 1 + \frac{2r^2}{r_0^2})\hbar$ [36], where $l = 1$. Figure 2(b) shows the dependence of the laser intensity and average angular momentum per laser photon on the normalized radial variable $r/r_0$. The peak intensity of the CP LG$_{10}$ laser is located at $r_{\max}/r_0 = \frac{\sqrt{2}}{2}$, where the gamma-ray photons are mostly generated. However, here, the average angular momentum per laser photon is only $1.0\hbar$. Considering the fact that a part of the CP LG$_{10}$ laser within $r < r_{\max}$ also contributes to the gamma-ray generation, it is reasonable that the OAM of the generated gamma-ray photons is approximate to $0.90\hbar$, which confirms the good agreement with the simulation result.

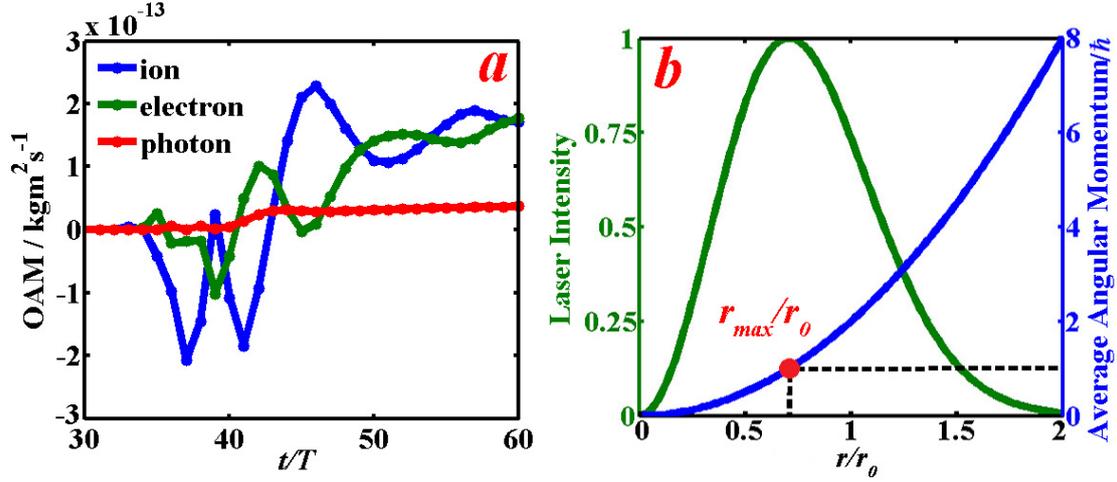

Fig. 2 (a) Variation of the OAM of the electrons (green line), protons (blue line), and photons (red line) with respect to time. (b) Dependence of the laser intensity (green line) and the average angular momentum per laser photon (blue line) on the normalized radial variable $r/r_0$. The red point indicates position of the peak intensity of the laser.

In addition, the transverse density of the gamma-ray photons at the rear target surface ($x = 31\lambda$) in the y-z plane also indicates that the gamma-ray photons are vortex. One snapshot of the transverse density at $t = 35.5T$ is shown in Fig. 3(a). Compared with snapshots of different times, it is found that the density structure of gamma-ray photons rotates around the x-axis. At the same position, the accumulated energy density of the gamma-ray photons integrated over the whole simulation time is shown in figure 3(b) and figure 3(c), where the CEP of the driving laser is $-0.5\pi$ and zero, respectively. It is not completely annular because the laser is ultra-short with only several periods. Therefore, the CEP of the laser is important for the photon energy density distribution and a relative change of the CEP $\Delta\phi_{CEP}$ leads to the relative change of the rotational angle $\Delta\phi_r = \Delta\phi_{CEP}/(l+s)$ in the pattern. Figure 3(d) shows the energy spectrum of the gamma-ray photons with a temperature of 10.75 MeV. Here, only photons with energy above 1 MeV are selected. Owing to the quantum radiation, the gamma-ray is incoherent and behaves more like a particle beam, which gives rise to a Boltzmann distribution of the energy spectrum.

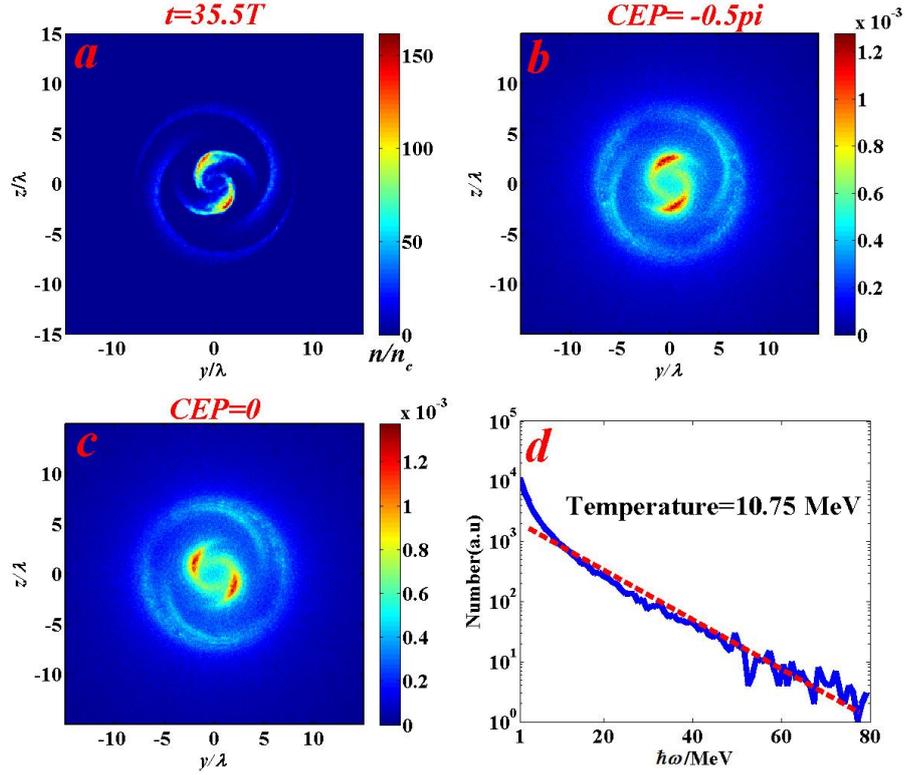

Fig. 3 (a) Density distribution of the gamma-ray photons at $x=31\lambda$ at $t=35.5T$ in the *y-z* plane. Accumulated energy density of the gamma-ray photons generated by the laser with (b) CEP=$-0.5\pi$, (c) CEP=0. (d) Energy spectrum of the gamma-ray photons. The red dashed line shows the data fitting of its temperature.

For high-mode CP LG lasers, the pattern of the generated gamma-ray photons is unique. The total energy density of the gamma-ray photons at $x=31\lambda$ is regularly divided into several parts, as shown in Fig. 4. The number of parts is equal to the angular momentum carried by each photon of the driving laser normalized to $\hbar$. In other words, it is equal to $l+s$. For example, for the right-hand CP LG$_{40}$ laser ($l=4, s=1$), the energy density of the gamma-ray photons consists of five parts; while for the left-hand CP LG$_{40}$ laser ($l=4, s=-1$), it consists of three parts. The reason is, for right-hand and left-hand CP LG lasers, the $E_y$ is the same while the $E_z$ has a phase difference of $\pi$. The phase difference in $E_z$ contributes to the different instantaneous intensity distributions of the laser and implies in the energy density of gamma-ray photons. The different intensity distributions of the right-hand and left-hand CP LG$_{40}$ lasers at

$x = 31\lambda$ is attached in the supplemental materials. It is found that the unique distribution of the energy density don't exist if a linearly polarized (LP) LG is used. This is because of the fact that the CP LG laser can provide a steady ponderomotive force and well-confine confines the electrons, while for the LP LG laser, there is an additional oscillating component of the ponderomotive force which hears the electrons quickly and causes them to scatter. Therefore, the topological charge and chirality of the driving CP LG laser is expected to be revealed according to energy density pattern. Given the laser spot $r_0$, by measuring the radical position $r_{max}$ where the gamma-ray photons are mostly generated, the topological charge $l$ can be known by $r_{max} = \frac{\sqrt{2l}}{2} r_0$. And the chirality ($s$) of the driving CP laser can be deduced based on the number $l+s$ of the pattern parts.

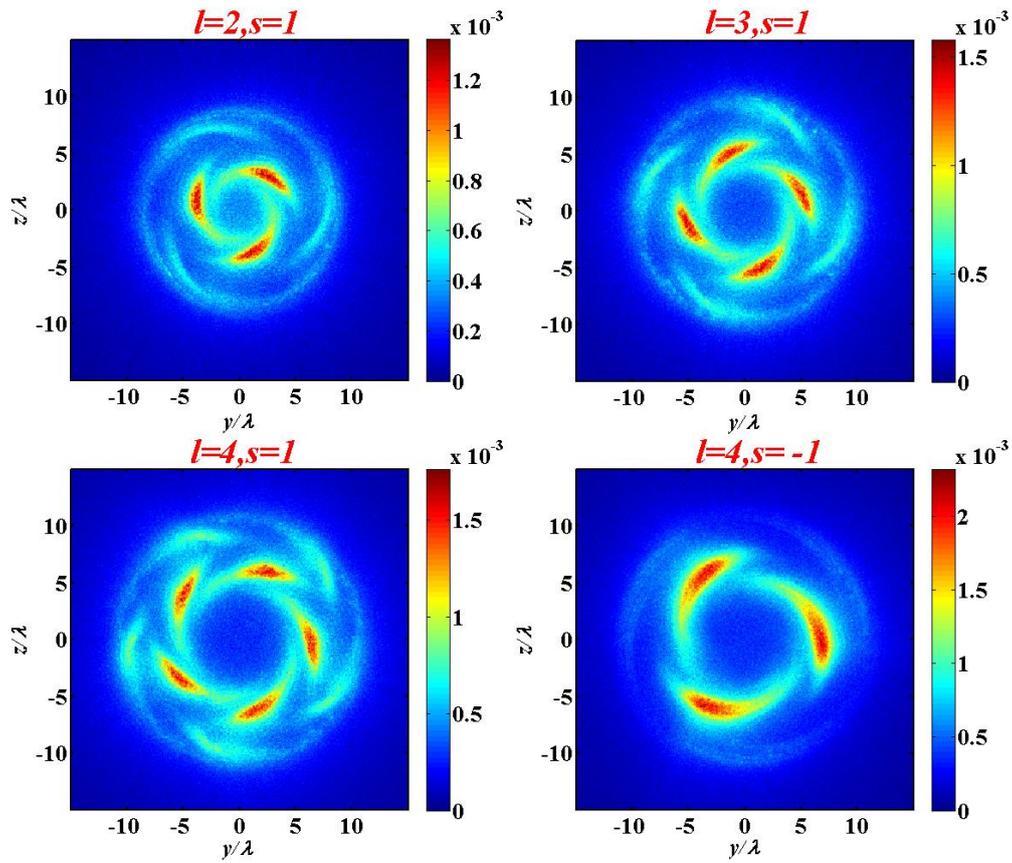

Fig. 4 Accumulated energy density of the gamma-ray photons at $x = 31\lambda$ for high-mode CP LG lasers with different topological charge $l$ and chirality $s$.

Furthermore, in order to determine what affects the OAM of the generated gamma-ray photons, two situations are considered. First, for a right-hand CP LG$_{10}$ laser, laser intensity improvement is certainly favorable for generating gamma-ray photons of higher OAM. In this series of simulations, a wide range of dimensionless laser amplitude $a_0$ has been investigated, varying from 100 to 1000, corresponding to $I_0$ from $4.28 \times 10^{22}$ W/cm$^2$ to $4.28 \times 10^{24}$ W/cm$^2$, with other parameters remain the same. Figure 5(a) shows that the OAM of the gamma-ray photons is nearly proportional to $a_0^2$. Second, when the topological charge $l$ of the LG laser increases, the laser pulse carries a higher OAM. Given the same laser energy, the OAM of different LG modes is linearly proportional to its topological charge $l$. In this series of simulations, the total energy of the laser is set to be the same, and only the topological charge changed from $l=1$ to $l=5$ (i.e., from LG$_{10}$ to LG$_{50}$). As expected, the dependence of the OAM of the gamma-ray photons on the topological charge $l$ is almost linear, as shown in Fig. 5(b). To summarize, the OAM of the generated gamma-ray photons is proportional to the total angular momentum of the driving vortex laser. For the laser with the same topological charge, the OAM of gamma-ray photons is proportional to the laser intensity; for the same energy of the CP LG laser, the OAM is proportional to the topological charges.

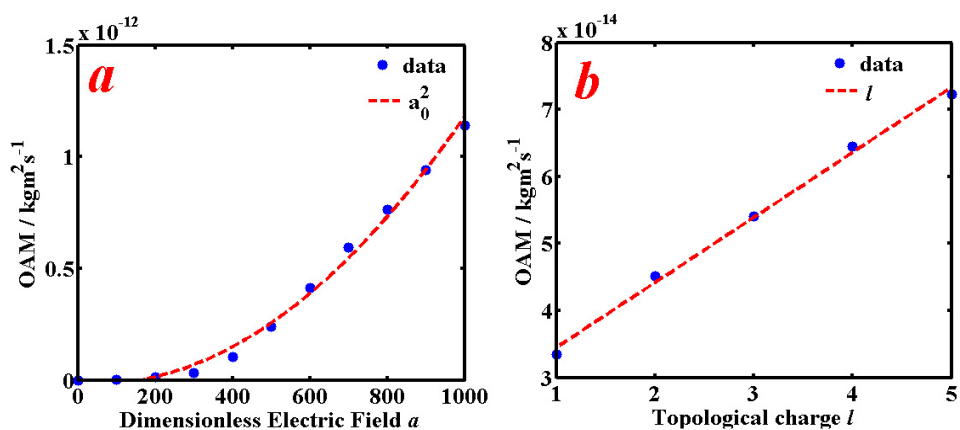

Fig. 5 Dependence of the OAM of the gamma-ray photons on the (a) dimensionless electric field $a_0$ and (b) topological charge $l$ of the LG laser. The red curve in

(a) shows the $a_0^2$ fitting of the OAM of the gamma-ray photons, and the red line in (b) presents the linear fitting of the topological charge.

In conclusion, a scheme to generate energetic gamma-ray photons with an OAM is proposed. It is found that by irradiating an intense CP LG$_{10}$ laser on a thin target, the SAM and OAM of the laser are transferred to electrons and then the OAM of electrons is transferred to gamma-ray photons through quantum radiation. Further studies revealed that the OAM of the gamma-ray photons is determined by the total angular momentum of the driving laser. To improve the OAM of the gamma-ray photons, two methods can be employed, either increasing the laser intensity or using a high-mode LG laser. For high-mode CP LG lasers, the pattern of energy density of the gamma-ray photons is unique. It consists of several parts and its number is equal to the angular momentum carried by each photon of the driving laser normalized to $\hbar$. And the CEP of the driving laser controls the relative rotational angle of this pattern. Therefore, the pattern of the accumulated energy density of gamma-ray photons is useful for detecting and revealing the topological charge, chirality and CEP of the ultra-intense vortex laser. The use of the ultra-thin target is to ensure that the laser pulse transmits the target completely so that a clear pattern of forward gamma-ray photons can be obtained. It also should be noted that the requirement of an ultra-thin target is not necessary. The gamma-ray beam with an OAM can also be achieved by using a target of wide thickness. However, for a thicker target, the reflection of the laser is enhanced and a portion of gamma-ray photons are generated backwards. In this situation, the accumulated energy density of the gamma-ray photons is less apparent than the ultra-thin target.

This work was supported by the National Natural Science Foundation of China (Grant Nos. 11335013, 61221064, 11374319 and 11127901).

**Supplemental Materials**

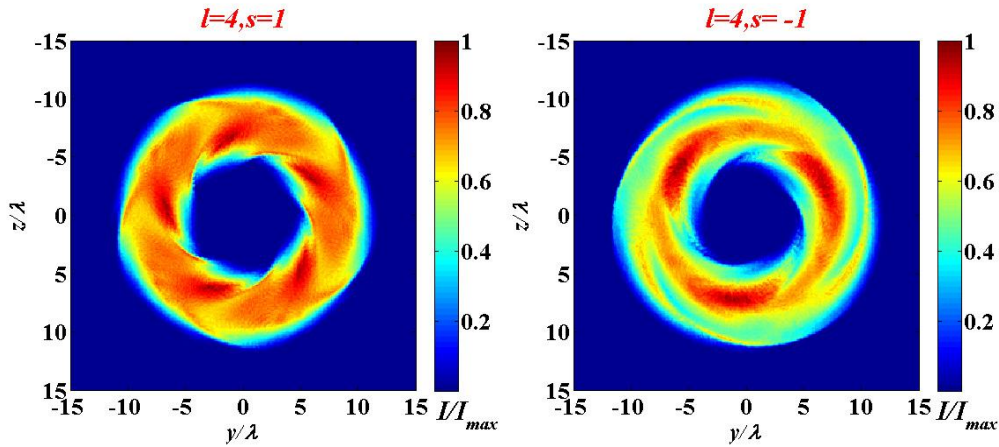

Snapshot of the different intensity distributions of the right-hand and left-hand CP LG$_{40}$ lasers at $x = 31\lambda$ during the laser-plasma interaction.